\documentclass[summary]{ursi}
\usepackage[nolist]{acronym}
\usepackage{hyperref}
\usepackage{amsmath,amssymb}
\usepackage{gensymb}
\usepackage[normalem]{ulem}
\usepackage{xcolor}
\usepackage{graphicx}
\usepackage{booktabs} 

\title{Radar Enabled Adaptive Modulation for Millimeter Wave Integrated Sensing and Communication}

\author{Jai Mangal, Sumit Darak, and Shobha Sundar Ram}
\affiliation{Indraprastha Institute of Information Technology Delhi, New Delhi 110020 India\\ (jaim,sumit,shobha)@iiitd.ac.in}
\begin{document}
\maketitle
\begin{acronym}
    \acro{BS}{base station}
    \acro{MU}{mobile user}
    \acro{RSP}{radar signal processing}
    \acro{BER}{bit error rate}
    \acro{ISAC}{integrated sensing and communication}
    \acro{mmWave}{millimeter wave}
    \acro{BPSK}{binary phase shift keying}
    \acro{QPSK}{quadrature phase shift keying}
    \acro{QAM}{quadrature amplitude modulation}
    \acro{OFDM}{orthogonal frequency division multiplexing}
\end{acronym}

\begin{abstract}
An integrated sensing and communication (ISAC) framework comprises radar sensing to enable reliable direction beam-based communication between a base station (BS) and mobile user (MU). The ISAC will be an integral part of 6G with potential applications for high-speed vehicular communications. Existing works have explored azimuth and Doppler velocity estimated via radar sensing for beam identification and identification in dynamic environments. In this work, we propose radar-enabled modulation scheme selection for ISAC, thereby eliminating conventional time-consuming downlink-uplink feedback-based modulation scheme selection. We have analyzed the performance of the proposed approach for four different trajectories and shown an improvement in throughput between 54-209\% over state-of-the-art ISAC. 
\end{abstract}

\section{Introduction}
Next-generation wireless systems demand high-speed data communication with ultra-low latency. One promising solution is to exploit the unlicensed bands within the \ac{mmWave} spectrum, which offers wide bandwidth and the opportunity to establish robust, high-capacity links essential for real-time data exchange \cite{chen2017v2x,alam2022integrated}. Due to substantial path loss at \ac{mmWave} frequency, narrow directional beam-based communication is needed, which demands quick identification of the appropriate beam between \ac{BS} and \ac{MU}. In the IEEE 802.11ad standard, such a beam is identified cooperatively by the BS and MU via beam refinement fields (BRF) embedded in the data frame. An integrated sensing and communication (ISAC) based modified 802.11ad is explored in \cite{sneh2024ieee,tewari2025reconfigurable} to address the long beam alignment time in standard 802.11ad due to iterative downlink and uplink communication overhead. The 802.11ad-based ISAC uses the Golay sequences within the packet preamble as a radar waveform and quickly estimates \ac{MU}'s range, azimuth, and Doppler velocity. The Doppler velocity estimate distinguishes the \ac{MU} from static clutter, while the azimuth estimate guides the subsequent directional communication. Other approaches, such as machine and deep learning, have also been explored in the literature \cite{ma2022deep,sneh2023radar,sneh2024beam} for faster beam identification. However, in all of these works, the assumption was that the modulation and communication schemes (MCS) remain fixed during the motion of the MU. Hence, even when the beams between the BS and MU are aligned, the communication throughput will fall when the link's signal-to-noise ratio (SNR) deteriorates due to various channel conditions. \\
\indent In addition to conventional downlink-uplink feedback-based modulation scheme selection, recent works have proposed adaptively switching between different modulation schemes based on the channel conditions. For example, \cite{hanawal_2024,neelamVLSI} explored multi-armed bandit (MAB) based online learning algorithms to quickly identify the optimal modulation scheme using SNR and interference information. Still, similar to beam alignment, selecting the appropriate scheme needs iterative cooperation between BS and MU over downlink and uplink, which results in a significant delay before communication with the optimal scheme is established. Such delay is usually long for high-speed MU, so a faster selection scheme is desired.\\
\indent To address this, we propose a modified ISAC framework in which the BS first estimates the range of the MU and then uses it to enable adaptive modulation schemes in the subsequent communication.
As the MU moves, the range between the \ac{BS} and \ac{MU} changes, based on which the MCS is adaptively switched between different schemes to maintain a high throughput and low bit error rate (BER). The proposed radar-enabled adaptive modulation-based ISAC is evaluated for multiple trajectories of \ac{MU}. We benchmark the performance of the proposed approach with standard 802.11ad-based ISAC with fixed modulation in \cite{sneh2024ieee} using two communication metrics - the throughput and average BER of the communication link. We have validated the superiority of the proposed scheme across four distinct MU trajectories and demonstrated average throughput improvements between 54-209\% over \cite{sneh2024ieee}.\\
\indent The paper is organized as follows. In Section II, we present the system model followed by the proposed work in Section III. The simulation results are presented in Section IV. Section V concludes the paper.
\section{System Model}
\begin{figure}[htbp]
  \centering
  \includegraphics[scale=0.2]{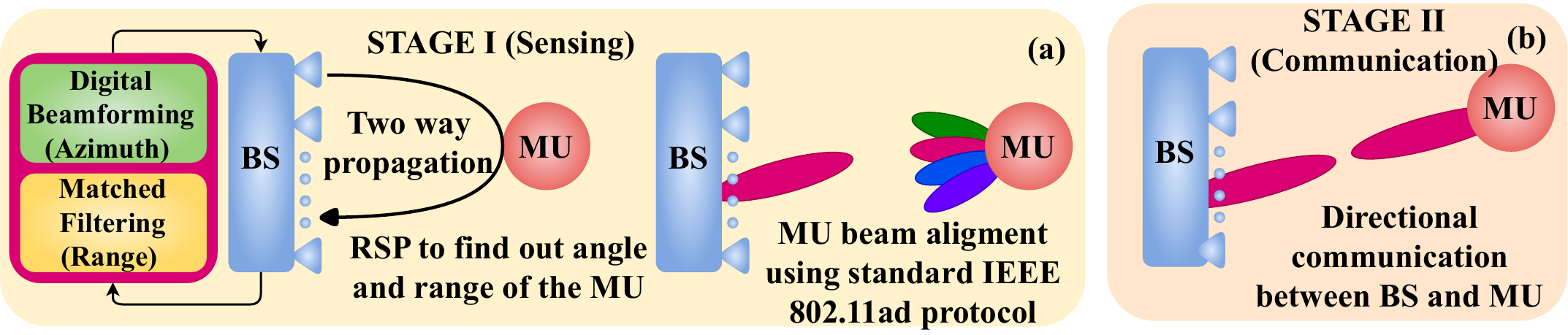}
  \caption{Modified IEEE 802.11ad based ISAC system. (a) Stage I (Sensing), (b) Stage II (Communication).}
  \label{fig:ISAC_stages}
\end{figure}

The standard IEEE 802.11ad packet, $\mathbf{y}$, comprises a preamble, header, data, and optional long BRF for facilitating beam alignment between the BS and MU \cite{ieee_2016-1}. In the modified ISAC-based 802.11ad ISAC system proposed in \cite{sneh2024ieee}, the packet preamble comprising Golay sequence, $\mathbf{x}[\tau]$, is exploited jointly for radar sensing and communication. Here, $\tau$ denotes the fast-time samples or bits within the waveform. Both functionalities then share the mm-wave hardware resources and spectrum in a time-division multiplex manner where the radar sensing forms the first stage, followed by communication. In this framework, we assume that the \ac{MU} for communication are mobile radar targets first detected and localized by the radar stage.
During the radar stage, $P$ radar waveforms, $\sum_{p=1}^P\mathbf{x}(t-pT_{PRI})$, are transmitted by BS omnidirectionally with a pulse repetition interval (PRI) of $T_{PRI}$. The radar signal is reflected by each $m^{th}$ MU modeled as a point target with a reflectivity of $a_m$ at a Doppler velocity, $v_m$, range $r_m$ at azimuth $\phi_m$. The BS receiver consists of an $N$ element uniform linear array antenna with $d_{BS}$ inter-element spacing that receives the three-dimensional (3D) reflected signal
\small
\begin{align}
\label{eq:RxSig}
\tilde{\mathbf{X}}[n,t,\tau] = \sum_{m=1}^Ma_m\mathbf{x}\left[\tau - \frac{2r_m}{c}\right]e^{-j2\pi f_mt}e^{-j(n-1)\frac{2\pi}{\lambda}d_{BS}\sin\phi_m}+\eta
\end{align}
\normalsize
for $t = 0 \cdots (P-1)T_{PRI}, \tau = 0 \cdots (Q-1)/F_s$ and $n=0 \cdots N-1$. In \eqref{eq:RxSig}, $\eta$ indicates the additive white Gaussian noise. $\tilde{\mathbf{X}}[n,t,\tau]$ is processed with matched filtering along fast time $\tau$ to estimate $r_m$, with Fourier processing along $t$ and $n$ to estimate $f_m$ and $\phi_m$ as shown in Fig.~\ref{fig:ISAC_stages}(a). 
In the subsequent communication stage, high-directional communication along the identified azimuth is established between BS and MU(s) as shown in Fig.~\ref{fig:ISAC_stages}(b). The MU comprises an $M$ element ULA with $d_{MU}$ inter-element spacing. The channel between the MU receiver and BS receiver is determined by the weight vectors at the BS, $\mathbf{w}_{BS} \in \mathbb{C}^{N \times 1}$ and MU, $\mathbf{w}_{MU} \in \mathbb{C}^{M\times 1}$, and the corresponding steering vectors - $\mathbf{u}_{BS} \in \mathbb{C}^{N \times 1}$ and $\mathbf{u}_{MU}\in \mathbb{C}^{M \times 1}$. The received communication signal, $\tilde{\mathbf{Y}}(t,\tau) \in \mathbb{C}^{M\times 1}$, is shown in
\begin{align}
\label{eq:CommRxSig}
\tilde{\mathbf{Y}}(t,\tau) = \mathbf{w}_{MU}\mathbf{u}_{MU}\mathbf{u}_{BS}^T\mathbf{w}_{BS}\mathbf{y}^T\left[\tau-\frac{r_m}{c}\right]e^{-j2\pi \frac{f_m}{2}t} + \zeta,
\end{align}
where the BRF fields are omitted from $\mathbf{y}$. In the above equation, $\zeta$ indicates the additive white Gaussian noise. 
The received downlink communication signal is downconverted, demodulated, and processed digitally to retrieve the data through a series of steps that include matched filtering, downsampling, timing synchronization, frame and preamble extraction for coarse and fine frequency offset estimation, cyclic prefix removal, channel estimation and equalization, phase synchronization, digital demodulation, and demapping. 
\section{Proposed Radar-Enabled Adaptive Modulation for ISAC}
\begin{figure}[!h]
  \centering
  \includegraphics[scale=0.28]{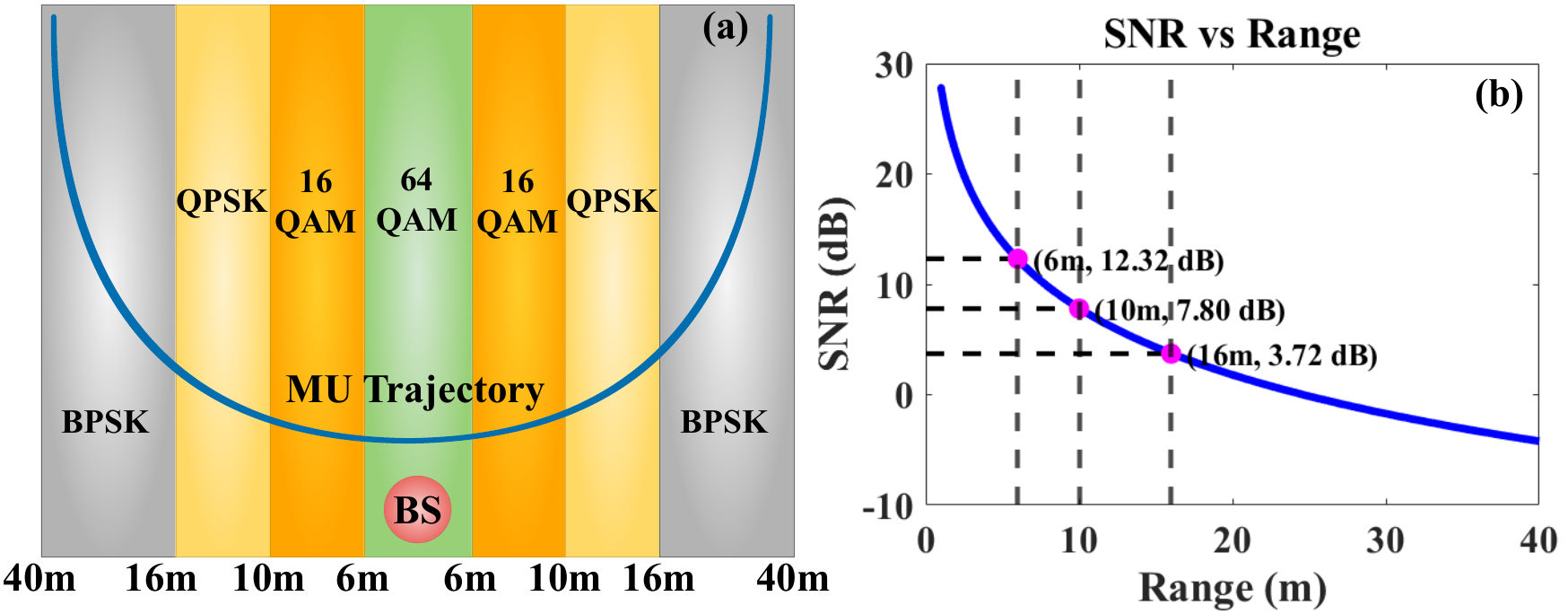}
  \caption{Radar enabled adaptive modulation. (a)Range-based proposed adaptive modulation scheme for ISAC, (b)Variation of SNR with increase in range.}
  \label{fig:snr_range}
\end{figure}
\begin{figure*}[!h]
  \centering
  \includegraphics[scale=0.48]{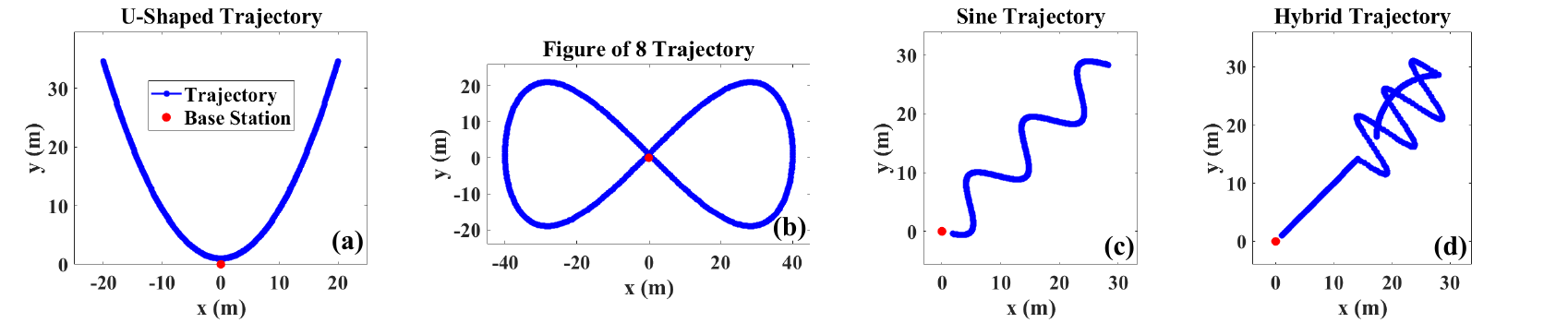}
  \caption{Trajectories for evaluation of adaptive modulation. (a) U-Shaped, (b) Figure of 8, (c) Sine and (d) Hybrid.}
  \label{fig:Trajectory}
\end{figure*}
\begin{figure*}[!h]
  \centering
  \includegraphics[scale=0.48]{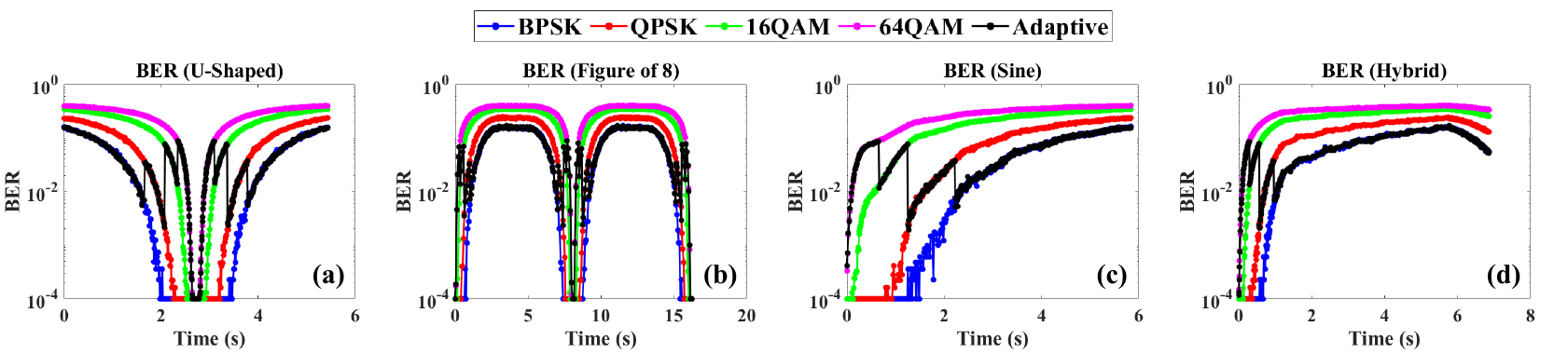}
  \caption{BER of ISAC system for different trajectories. (a) U-Shaped, (b) Figure of 8, (c) Sine and (d) Hybrid.}
  \label{fig:BER}
\end{figure*}
\begin{figure*}[!h]
  \centering
  \includegraphics[scale=0.48]{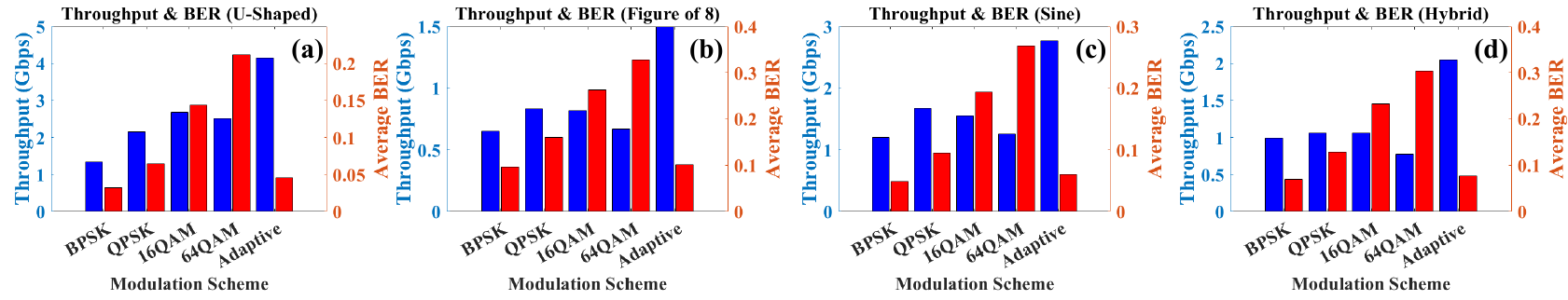}
  \caption{Throughput and average BER for different trajectories. (a) U-Shaped, (b) Figures of 8, (c) Sine and (d) Hybrid.}
  \label{fig:Throughput_BER}
\end{figure*}
This section presents the radar-based adaptive modulation scheme for the ISAC system. As discussed in the previous section, we perform sensing and communication in a time division duplexing manner in two stages. In stage I, we perform radar-based sensing. We transmit a 768-bit Golay sequence extracted from the IEEE 802.11ad preamble with 50\% duty cycle, omnidirectionally at the \ac{mmWave} frequency of 60 GHz. We have utilized the Golay sequence because of its perfect correlation property, resulting in a very high peak-to-side lobe level along the range.
The MU reflects the signal to the \ac{BS}. 
Once we figure out the range and angle of the \ac{MU} from \ac{BS}, we initialize stage II, where we transmit an \ac{OFDM} signal at 60 GHz directionally by steering the beam toward the \ac{MU} using analog beamforming. We transmitted 384 modulated complex samples per OFDM symbol, to which 64 NULL symbols were added on both ends, resulting in 512 complex symbols. The OFDM modulation comprises 512 points IFFT and 128 lengths of cyclic prefix addition are done. The digital modulation schemes used while transmitting the OFDM signal are BPSK, \ac{QPSK}, 16 \ac{QAM}, and 64 QAM. For BPSK, QPSK, 16-QAM, and 64-QAM modulation scheme, the number of transmit samples per symbol is 2, 4, 4, and 6, respectively. However, as the number of samples per symbol increases, the spacing between decision boundaries in the constellation diagram decreases. Consequently, at lower SNRs, where constellation points (i.e., IQ symbols) are more dispersed due to noise, the system can switch to a lower-order modulation scheme to maintain the BER. Conversely, when the constellation points are more tightly clustered around their ideal locations at higher SNR, higher-order modulation can be employed to achieve greater throughput while maintaining an acceptable BER. This work proposes an adaptive digital modulation scheme that exploits the range information between the \ac{BS} and \ac{MU} to determine the best digital modulation scheme to maximize throughput and minimize the BER of the system. The range at which the modulation scheme needs to be switched by the BS is determined based on the SNR at the MU receiver for that particular range. We estimate the SNR at the MU using \eqref{eq:snr} for a given range.

\begin{equation}
    SNR_{MU} = \frac{P_tG_tG_r\lambda^2}{(4\pi d)^2\ N_p}
    \label{eq:snr}
\end{equation}
where, $P_t$ is the transmit power, $G_t$ and $G_r$ are the gain of transmit and receive antennas, $d$ is the range between BS and MU, and $N_p$ is the noise power of the MU. Based on the estimation SNR, an appropriate modulation scheme is selected, as shown in Table~\ref{tab:adap_mod}. For the same value of SNR, the performance of the higher modulation scheme will be poorer than that of the lower modulation scheme. Because of this, the modulation scheme needs to be adapted based on the SNR values. However, the SNR depends on the range as shown in Fig.~\ref{fig:snr_range}(b). As the range increases, it will reduce received signal power due to path loss, and for a given noise power at the receiver, the SNR decreases as a function of the increase in range. In this work, the SNR cutoffs are mapped to the range cutoff as presented in Table~\ref {tab:adap_mod} using Figure~\ref {fig:snr_range}(b). As the MU travels through the trajectory, the BS switches the MCS based on the range presented in Fig.~\ref {fig:snr_range}(a). In stage I (sensing), the BS determines the range of MU from the BS and utilizes this to switch the MCS accordingly for stage II (communication).
\begin{table}[htbp]
\centering
\caption{SNR mapped range-based adaptive MCS}
\label{tab:adap_mod}
\begin{tabular}{ccc}
\toprule
\textbf{Range (m)} & \textbf{SNR (dB)}       & \textbf{Modulation Scheme} \\ 
\midrule
$< \ 6$              & $>12.32$                & 64-QAM                     \\[0.5ex]
$6\ -\ 10$        & $12.32\ -\ 7.80$         & 16-QAM                     \\[0.5ex]
$10\ -\ 16$       & $7.80\ -\ 3.73$          & QPSK                       \\[0.5ex]
$> \ 16$            & $< \ 3.73$                & BPSK                       \\ 
\bottomrule
\end{tabular}
\end{table}
\section{Performance Analysis}
In this section, we compare the performance of the proposed radar-enabled adaptive modulation based \ac{ISAC} and the benchmark ISAC in \cite{sneh2024ieee} for four trajectories of the MU as shown in Fig.~\ref{fig:Trajectory}. The BS and MU parameters used in our simulations are given in Table~\ref {tab:system_params}. We consider a ground area of $80m\times40m$, in which the \ac{BS} is stationed at the origin and the \ac{MU} moves at a constant velocity of $15m/s$. We use the BER, calculated for each data frame, and the throughput as the communication performance metrics. The throughput is calculated as
\begin{equation}
\text{Throughput} = (F_{s}) (overhead) (\log_2 M) (1-\overline{BER})S
\label{eq:throughput}
\end{equation}
where $(F_{s})$ is the symbol or sampling rate, which is 2.64 Gbps, overhead is $(N_{Data})/(N_{FFT} + N_{CP})$ which is 0.6, $M$ is the modulation order which is 2 for BPSK, 4 for QPSK and 16QAM, and 6 for 64 QAM. The flange $S$ indicates successful reception of the frame, which is assumed when the average BER is below 0.1.
\begin{table}[ht]
  \centering
  \caption{BS and MU system parameters.}
  \label{tab:system_params}
  \begin{tabular}{cc}
    \toprule
    \textbf{System Parameter} & \textbf{Value} \\
    \midrule
    PRI & 0.58 $\mu s$ \\
    Duty Cycle & 50 \% \\
    Maximum Unambiguous Range & 43.5m \\
    No. of Antennas at BS \& MU & 16 \& 4 \\
    RCS of Point Target & 0 dBsm \\
    Noise Power of MU & 10pW (-80 dBm) \\
    \bottomrule
  \end{tabular}
\end{table}
We begin with the U-shaped trajectory of the MU, as shown in Fig.~\ref{fig:Trajectory}(a), In this trajectory, MU starts at a distance of 40m from the BS, reaches near the BS after 3 seconds and moves away from the BS thereafter. As shown in Fig.~\ref{fig:BER}(a), the proposed adaptive ISAC sequentially switches from lower to higher modulation schemes in the first 3 seconds and from higher to lower modulation schemes in the next 3 seconds. Overall, the proposed scheme for a U-shaped trajectory achieves average BER and throughput of 0.045 and 4.15 Gbps, as shown in Fig.~\ref{fig:Throughput_BER}(a). When compared to fixed-modulation-based ISAC, the adaptive MCS provides throughput improvement by 209\%, 92\%, 54\% and 65\% for BPSK, QPSK, 16-QAM and 64-QAM respectively.\\
We next consider the figure-of-8 trajectory as shown in Fig.~\ref{fig:Trajectory}(b). The MU starts 1 m from the BS and moves away for 3 seconds, triggering a switch to lower-order modulation due to reduced SNR. Over the next 2 seconds, the MU maintains a 40 m distance before moving back toward the BS over 3 seconds, prompting a return to higher-order modulation. This pattern repeats over the remainder of the trajectory as shown in Fig.~\ref{fig:BER}(b). Overall, the proposed range-enabled ISAC achieves 1.50 Gbps throughput and a 0.10 average BER as shown in Fig.~\ref{fig:Throughput_BER}(b), with throughput improvements of 130\%, 80\%, 83\%, and 123\% over BPSK, QPSK, 16-QAM, and 64-QAM, respectively. Similarly, for sine trajectory in Fig.~\ref{fig:Trajectory}(c), an improvement of 130\%, 65\%, 79\%, and 119\% over BPSK, QPSK, 16-QAM and 64-QAM, respectively is realized. For the hybrid trajectory, shown in Fig.~\ref{fig:Trajectory}(d), an improvement of 107\%, 93\%, 93\%, and 162\% over BPSK, QPSK, 16-QAM, and 64-QAM, respectively, is observed. This demonstrates the advantages of a radar-enabled adaptive modulation scheme over fixed modulation schemes.
\section{Conclusion}
This paper proposes a radar-enabled adaptive modulation scheme based on ISAC. In the ISAC framework, radar sensing detects and localizes MU, which helps improve communication performance via accurate beamforming. In the literature, azimuth and Doppler velocity estimates obtained via radar sensing are used. This work further enhances it by incorporating range information to switch between different modulation schemes, leading to faster identification of optimal modulation schemes than conventional downlink-uplink feedback-based selection. We have validated the superiority of the proposed scheme across four distinct trajectories, demonstrating average throughput improvements between 54-209\% over state-of-the-art ISAC.

\bibliographystyle{ieeetr}
\bibliography{ref_new}

\end{document}